\documentclass[a4paper,12pt]{revtex4-1}

\usepackage{fullpage}
\usepackage{graphicx}
\usepackage{subfigure}
\usepackage{epsfig}
\usepackage{dcolumn}
\usepackage{bm}
\usepackage{amsmath,amssymb,color}
\usepackage{setspace}

\begin{document}

\title{Second law, Landauer's Principle and Autonomous information machine}

\author{Shubhashis Rana$^{1}$ }
\email{ shubhashis.rana@gmail.com }
\author{A. M. Jayannavar$^{2,3}$ }
\email{ jayan@iopb.res.in}
\affiliation{ $^1$S. N. Bose National Centre for Basic Sciences JD Block, Sector-III, Salt Lake City, 
Kolkata - 700 106, India. \\ $^2$Institute of Physics, Sachivalaya Marg, Bhubaneswar 751005, India.
\\$^3$Homi Bhabha National Institute, Training School Complex, Anushakti Nagar, Mumbai 400085, India. }



\begin{abstract}
Second law of thermodynamics can be apparently violated for systems whose dynamics depends on  acquired information  by 
measurement. However, when one consider measurement and erasure process together along with the system it saves the second
law. 
  We consider a simple example of information machine where information is used as a resource to increase
 its performance. The system is connected to 
 two baths, a work source and a moving  tape which is used as an information reservoir. The performance of the
 device is autonomous. The system acts as an engine, erasure or refrigerator. Even combination of any
 two is possible. All these possibilities are allowed by generalized second law.
  \end{abstract}

\maketitle{}

\newcommand{\nwc}{\newcommand}
\nwc{\vs}{\vspace}
\nwc{\hs}{\hspace}
\nwc{\la}{\langle}
\nwc{\ra}{\rangle}
\nwc{\lw}{\linewidth}
\nwc{\nn}{\nonumber}
\nwc{\tb}{\textbf}
\nwc{\td}{\tilde}
\nwc{\Tr}{\tb{Tr}}
\nwc{\dg}{\dagger}

\nwc{\pd}[2]{\frac{\partial #1}{\partial #2}}
\nwc{\zprl}[3]{Phys. Rev. Lett. ~{\bf #1},~#2~(#3)}
\nwc{\zpre}[3]{Phys. Rev. E ~{\bf #1},~#2~(#3)}
\nwc{\zpra}[3]{Phys. Rev. A ~{\bf #1},~#2~(#3)}
\nwc{\zjsm}[3]{J. Stat. Mech. ~{\bf #1},~#2~(#3)}
\nwc{\zepjb}[3]{Eur. Phys. J. B ~{\bf #1},~#2~(#3)}
\nwc{\zrmp}[3]{Rev. Mod. Phys. ~{\bf #1},~#2~(#3)}
\nwc{\zepl}[3]{Europhys. Lett. ~{\bf #1},~#2~(#3)}
\nwc{\zjsp}[3]{J. Stat. Phys. ~{\bf #1},~#2~(#3)}
\nwc{\zptps}[3]{Prog. Theor. Phys. Suppl. ~{\bf #1},~#2~(#3)}
\nwc{\zpt}[3]{Physics Today ~{\bf #1},~#2~(#3)}
\nwc{\zap}[3]{Adv. Phys. ~{\bf #1},~#2~(#3)}
\nwc{\zjpcm}[3]{J. Phys. Condens. Matter ~{\bf #1},~#2~(#3)}
\nwc{\zjpa}[3]{J. Phys. A: Math theor  ~{\bf #1},~#2~(#3)}

\section{Introduction}
  
Second law of thermodynamics had been and will always be one of the strongest rules of nature.
However, its validity has been questioned many times but it always hold on average.
Challenging the validity of the second law dates back nearly 150 years ago, when 
Maxwell proposed his famous thought experiment \cite{max71}.  
In his gedanken  experiment he considered  a box filled with  gas. Now, this box is divided  into two parts
by a partition.  In that partition there is a small door whose dimension is comparable
to the gas molecule.  A demon monitors the gas particles and allow to pass them through the door.
When a faster molecule comes  from left, the demon allows it  to transfer into
right side by opening the door, while the demon closes the door if a slower molecule comes from 
the left. On the other hand, the demon only allows to  transfer slower molecules from right 
side to the left.  Hence over a  course of time sorting takes place and right part of the box
contains faster molecules compared to the left. Since  the average kinetic  energy of 
gas particles determines the temperature of the gas, over time colder side become more colder and hotter side 
become more hotter. This happens when the  demon monitors the  gas particles by only knowing velocity of 
each individual gas molecules.  This seems to be violation of the second law.
Because the second law forbids  transfer of heat from cold body to hot body without doing any work.
 
 There is another famous example, given by Szilard in 1921, where a demon is involved and second law
 is apparently broken \cite{szi29}. He formulated an engine that can extract energy from 
 a single heat bath and convert it into useful work in presence of information in a cyclic process. 
 In his original derivation he took a single particle confined in a 
box of volume V and the box is connected to a bath with temperature T. A demon is placed to monitor the system and
extract energy. First the demon put a partition quickly in the middle and separate the box into two equal parts. The 
gas molecule would be confined into any one of the parts. Then it measures in which part the molecule is in.
Depending on the measurement, the partition is moved isothermally such that $k_B T \ln2$ amount of work is extracted. Finally
the partition is removed and the system goes to its original state. For classical systems, insertion and removal of the 
partition do not need any energy. Hence, one can extract energy from heat bath repeatedly  which is surprising
and the system is termed as a perpetual machine. However, it is unfair to treat a single particle kicking randomly to the 
wall of the box as an ideal gas. But, any system that undergoes a phase-space splitting can be used 
as a working system to harness energy. In both the cases the demon uses its acquired information 
smartly and decreases the entropy of the system without performing any work by himself. 
 It can be shown that one can perform measurement without involving any energy cost. The paradox
of this  apparent violation of second law can be resolved if one includes memory of the 
demon as a part of the whole system. The memory which records the state of the system has to be
reset to its initial state. This process is called 'erasure'. It is a irreversible process because
there is no one to one correspondence between input and output of this process. According to 
the Landauer principle any logically irreversible transformation of classical information 
is necessarily accompanied by dissipation of at least $k_B T ln 2$ of heat per lost bit \cite{lan61}.
This leads to an entropy increase of the bath at least by $k_B ln 2$ per bit. Entropy cost for resetting demon's memory 
is always larger than the initial entropy reduction, thereby safeguarding the second law.
 The importance of Szilard Engine and Maxwell demon is that they made 
information in the same footing as entropy that will be discussed below in more detail. 

\section{The  second law in non-equilibrium regime}

In information theory, the Shannon entropy of a random variable X with probability density at any time  $\rho (x,t)$,
is defined as $\mathcal{H}(X,t)=-\Tr \rho(x,t) \ln \rho(x,t)$. Shannon entropy  denotes the uncertainty of the random variable and
measures the amount of information required to describe the random variable.  When X represents the microscopic state of a physical system  
one can define the non-equilibrium  entropy as
\begin{equation}
 S(t)=k_B \mathcal{H}(X,t)=-k_B\Tr \rho(x,t) \ln \rho(x,t),
 \label{noneq-ent}
\end{equation}
where $k_B$ is the Boltzmann constant. Recent development of stochastic thermodynamics reveals that the Shannon
entropy has a clear meaning in certain situation and  determines the energetics in non-equilibrium processes
for systems connected to one or more thermodynamic reservoirs. 
Let us consider a system connected to a bath of temperature T.  The system is evolved by a protocol $\lambda(t)$
and  $H(\lambda(t))$ denotes corresponding  time dependent Hamiltonian. Then the system energy at any time is given by

\begin{equation}
 E(t)=\Tr \rho(t)H(t).
\end{equation}
Corresponding non-equilibrium free energy is defined as \cite{esp11}

\begin{equation}
\mathcal{ F}(t)=E(t)-TS(t).
\end{equation}
Here T denotes the temperature of the bath to which the system is connected. Suppose the protocol is kept fixed at a particular
value of  $\lambda(t)$ then the  system will reach to the equilibrium distribution
\begin{equation}
 \rho ^{eq}(t)=\frac{1}{Z(\lambda(t))}\exp^{-\beta H(\lambda(t))},
\end{equation}
with $\beta=1/k_B T$ the inverse  bath temperature and the partition function $Z(\lambda(t))=\Tr \exp^{-\beta H(\lambda(t))}$.
Note that  equilibrium distribution should not depend on time. Here the time in the bracket just denotes the 
value of the protocol $\lambda(t)$ at which the equilibration is done. Now if we put $\rho^{eq}(t)$ in  
Eq.(\ref{noneq-ent}) then the non-equilibrium entropy will coincide with the corresponding equilibrium entropy
and one can recover usual relation $F(t)=\Tr \rho^{eq}(t)H(t)-TS$ with $F(t)$ as equilibrium free energy when the
protocol fixed at the value of $\lambda(t)$ 
Suppose the system is initially in a non-equilibrium state $\rho(0)$. Then it is evolved by an external 
 protocol $\lambda(t)$   upto time $\tau$ and the system reaches to another non-equilibrium state $\rho(\tau)$.
$W$ and $Q$ represents the work done on the system and heat transferred to the bath during this process. 
Then according to the first law of thermodynamics, the internal energy change of the system 
during this evolution becomes
\begin{equation}
 \Delta E=W- Q,
\end{equation}
The change in non-equilibrium system entropy
\begin{equation}
 \Delta S=\Delta S_{tot} - \Delta S_B
\end{equation}
consists of two terms. Total entropy production $\Delta S_{tot}\ge 0$ is strictly positive quantity.
 The second term is the entropy flow to the bath and is defined as $\Delta S_B=Q/T$.
Using first law and the definition of non-equilibrium free energy one can get the Second law for non-equilibrium regime
\begin{equation}
 T\Delta S_{tot}=W-\Delta \mathcal{F}\ge 0.
\end{equation}
Where $\Delta \mathcal{F}= \mathcal{F}(\tau)-\mathcal{F}(0)$.  That means the extractable work $-W$ is always bounded by decrease of non-equilibrium free energy 
difference $-\Delta \mathcal{F}$. This is a more general equation. Now if the system starts from equilibrium and reaches 
to another equilibrium point after evolution then the above equation will be reduced to the usual equation 
\begin{equation}
 W_{diss}=W-\Delta F\ge 0,
\end{equation}
where $W_{diss}$ denotes dissipated work and  $\Delta F= F(\tau)-F(0)$ is the equilibrium free energy change.

\section{Relative entropy and Mutual Information}
In Information theory, the relative entropy or Kullback Leibler distance measures the distance between two distributions and is defined as 
\begin{equation}
 D(p||q)=\sum_x p(x)\ln\frac{p(x)}{q(x)}.
\end{equation}
This distance the distinguishability between two distributions. 
However, it is not symmetric.  If the two distributions
coincide, $p(x)=q(x)$ then  relative entropy vanishes. Physically  $\mathcal{H}(p)+D(p||q)$ is the number
of bits required on average to describe the random variable in terms of $q(x)$.
 Using the property, $\ln(x)\le x-1$ for $x\ge 0$ we have
\begin{eqnarray}
  D(p||q)&&=\sum_x p(x)\ln\frac{p(x)}{q(x)}\nn\\
         &&=-\sum_x p(x)\ln\frac{q(x)}{p(x)}\nn\\
&& \ge \sum_x p(x)\left[1-\frac{q(x)}{p(x)}\right]=\sum_x [p(x)-q(x)]=1-1=0,\nn\\
\end{eqnarray}
which implies that relative entropy is always positive.
 The mutual information between two random variables U and V  is defined as
\begin{equation}
 I(U;V)=\sum_{u,v}\rho(u,v)\ln\frac{\rho(u,v)}{\rho(u)\rho(v)}=\mathcal{H}(U)+\mathcal{H}(V)-\mathcal{H}(U,V).
\end{equation}
Mutual information is basically the relative entropy between the joint distribution of the two random variables with
their product distribution. Note that, unlike relative entropy,  mutual information is symmetric i.e, $I(U;V)=I(V;U)$.
 $I(U;V)$ is always positive and vanishes only when these two random variables U and V are statistically independent i.e.,
there is exists no correlation between them. One can rewrite  $I(U;V)$ as 
\begin{equation}
  I(U;V)=\mathcal{H}(U)-\mathcal{H}(U|V)=\mathcal{H}(V)-\mathcal{H}(V|U),
\end{equation}
where $\mathcal{H}(U|V)=-\sum_{u,v}\rho(v)\rho(u|v)\ln\rho(u|v)$ denotes the conditional entropy and quantifies uncertainty 
of U for given V. Now, $\mathcal{H}(U)$ represents uncertainty of U. Hence, 
mutual information denotes the reduction of uncertainty of one random variable due to knowledge of another. 


\section{Measurement and Entropy}

We will now discuss the effect of 
measurement on a system with  probability density   $\rho(x)$. Suppose, a measurement is performed and  m is the outcome
like left or right in the Szilard engine. The probability density will be updated to $\rho(x|m)$ after this measurement.
 If the system is initially in equilibrium, $\rho(x|m)$ will not be canonical  due to measurement. 
Thus measurement leads the system to a non-equilibrium state although there is no energy cost. The change in non-equilibrium
entropy due to measurement is $S(\rho(x|m)-S(\rho(x))$. Here  $ S(\rho(x|m))=-k_B \sum_x \rho(x|m) \ln \rho(x|m)$ and 
$ S(\rho(x))=-k_B \sum_x \rho(x) \ln \rho(x)$.
Taking the average over all possible outcomes with probability
$p_m$, the non-equilibrium entropy change becomes
\begin{eqnarray}
 \Delta S_{meas}&&=-k_B \sum_{x,m} p_m \rho(x|m) \ln \rho(x|m)+ k_B \sum_{x.m} p_m \rho(x) \ln \rho(x)\nn\\
 &&=k_B(\mathcal{H}(X|M)-\mathcal{H}(X))\nn\\
 &&=-k_B I(X;M).
\end{eqnarray}
Note that, in measurement process neither the Hamiltonian nor the micro-state of the system is affected. That means the average energy 
of the system does not change due to measurement. Hence the non-equilibrium free energy change will become 
\begin{equation}
 \Delta \mathcal{F}_{meas}=\sum_m p_m\mathcal{F}(\rho(x|m);\mathcal{H})-\mathcal{F}(\rho(x);\mathcal{H})=-T\Delta S_{meas}=k_BT I(X;M).
\end{equation}
As, $ I(X;M)\geq 0$ there is always increase in non-equilibrium free energy which can be eventually used to extract work 
in an isothermal process at later times. This explains the functioning of a Szilard engine in a cyclic process.

\section{Landauer Principle and memory}

Information seems to be an abstract quantity at first glance. However, it is not. When a measurement is performed,
the obtained information is being stored in a piece of paper  or in the hard-disk etc.
 In these perspective, Landauer shed new light in his famous article 
'Information is physical' \cite{lan91}. Any physical system with multiple distinguishable
meta-stable states can be used  to store the information. However, these  states should have 
long enough lifetimes and should not be affected by the   environmental fluctuations or any external constraints. Then only
a system can act reliably as memory in desired time.  It means ergodicity must be broken or effectively broken in 
the timescale when memory is reliable. The total phase space $\Gamma$ is split into several ergodic regions
 $\Gamma_{m}$ for each memory outcome $m$. Magnetization in a small ferromagnetic domain of a standard magnetic memory or high
energy barriers separating microscopic degrees of freedom in single electron memory are few examples in this regard.
 
Let $p_m$ denotes the probability to be in the ergodic region $\Gamma_m$ of the memory. Now if the memory is in local 
equilibrium, one can take $E_m$ and $S_m$ as average energy and non-equilibrium entropy of the corresponding 
ergodic region. The non-equilibrium free energy of the memory can be written as \cite{par15}
\begin{eqnarray}
 \mathcal{F}(M)=\sum_m p_m F_m-k_BT\mathcal{H}(M),
\end{eqnarray}
where, $F_m=E_m-TS_m$ and $\mathcal{H}(M)=-\sum_m p_m\ln p_m$ is the Shannon entropy of the informational states.
Note that, total entropy of the memory is sum of Shannon entropy $\mathcal{H}(M)$  and
 the individual internal entropies $S_m$.  
Now after manipulation of the memory, we assume that the Hamiltonian of the system reaches to its initial Hamiltonian.
Then we need to concern only about the expression of $p_m$ for particular memory state. Suppose during measurement 
the state changes from $M'$ with probability $p_m'$ to the state $M$ with probability $p_m$. The change in free
energy for the memory for this case will be $\Delta\mathcal{F}^{mem}_{meas}=\mathcal{F}(M)-\mathcal{F}(M')$.
Similarly for resetting the memory to its original state $M'$ the free energy change will be
 $\Delta\mathcal{F}^{mem}_{reset}=\mathcal{F}(M')-\mathcal{F}(M)$ and work done to reset the memory must follow 
\begin{equation}
  W_{reset}\ge \Delta\mathcal{F}^{mem}_{reset}.
 \end{equation}
In this respect, if one takes symmetric memory $F_0=F_1=F_2=...$, the free energy change will be reduced to 
only the Shannon entropy change $\Delta\mathcal{F}^{mem}_{reset}=-k_BT(\mathcal{H}(M')-\mathcal{H}(M))$.
Now if we reset the state to a standard state such that $p'_0=1$ and all other $p'_m=0$ then $\mathcal{H}(M')=0$.
This is the ``restore to zero process'' and one gets   $\Delta\mathcal{F}^{mem}_{reset}=k_BT\mathcal{H}(M)$.
If the memory consist of only two 
states and for random bit $p_0=p_1=1/2$, one obtain the celebrated Landauer's limit 
\begin{equation}
 W_{reset}\geq k_BT\ln2,
\end{equation}
while equality holds for reversible processes. This is the famous Landauer's principle which states that the 
 minimum work need to erase one bit of information is $ k_BT\ln2$. The opposite of the restore to zero 
 processes increases  disorder in the states of memory, as a result, one can easily extract work out of this process.
   This trick is used in recently proposed autonomous information
 engine model where information is used as a fuel to extract thermodynamic 
 work in isothermal process.  Hence an ordered state of memory registrar can be treated as a reservoir of information
 and can be taken to be in same footing with other thermodynamic reservoirs like thermal or chemical bath.

\section{An Autonomous Information Machine: Working As Engine,
Refrigerator And Erasure}

In this section we have described a model where information is used as a fuel to extract work \cite{ran16}.
 The model consists of a system (demon), a work source and an information reservoir (Figure 2). The system 
 can interact with two thermal reservoirs, hot and cold.   A mass that can be lowered by gravity or pulled up against 
gravity is used as a work source. A stream of bits written on a tape is used as an information reservoir.
The information can be written/ removed  during the operation. 
The system can exchange information with the information reservoir but 
not energy.  Although, it can exchange energy with the work source and the thermal bath.  
The dynamics is taken autonomous that means there is no external control. 
The system reaches unique steady state depending on its parameter. The main 
motivation of this study is to find the effect of thermal bias along with the work source and
informational source. Recently D Mandal And C Jarzynski has given a simple  model \cite{man12} where the system 
is connected with single bath and it exhibit erasure and engine. 
In another model \cite{man13} it exhibit refrigerator and erasure. In the present model one can get engine, 
refrigerator and erasure simultaneously in a single model.  The system can perform as 
engine by extracting energy out of the thermal reservoirs, refrigerator by transferring heat
from cold to hot bath and erasure by removing information content in the tape. We will explain
these definitions in more detail in later section. It is observed, even more surprisingly, 
the system can perform any two of these behaviors simultaneously! However, these behaviors are
consistent with the second law of thermodynamics. Moreover,  the efficiency of the engine
and the coefficient of performance of the refrigerator can go beyond the Carnot limit.

\subsection{The Model}
Let us take a three state system (A, B, C) and these states are non-degenerate Fig.(\ref{model2}).
The difference between the two successive energy levels is taken to be same ($E_1$). 
The transition can occur between A to B, B to C and vice versa by exchanging heat from the
cold bath spontaneously with temperature $T_c$ and internal energy of the system gets changed.
However, the transition between A to C and vice versa is restricted. It depends on the value 
of the interacting bit written on the tape. As the bit has two states 0 and 1, 
the combined system has 6 states.   The bit state is changed from 0 to 1 when the transition
occur from C0 to A1 ( clockwise  rotation ) and  vice versa ( anti-clockwise  rotation ).
However, transition between C1 and A0 is not allowed.  Note that, the bit state do not change
when transition occurs between A and B, B and C. During the transition  from C0 to A1 (Clockwise  rotation), 
energy is absorbed from the hot bath at temperature $T_h$ by an amount $E$ while the system performs
w amount of work by pulling a mass m to a height $\Delta h$ by a frictionless pulley
in  a gravitational force field g ($w = mg \Delta h$). Note that, during this transition, 
the internal energy of the demon is increased by $2E_1$. Using first law one can write

\begin{equation}
 E=w+E_1.
\end{equation}
\begin{figure}[!ht]
\vspace{0.5cm}
\begin{center}

\includegraphics[width=7cm]{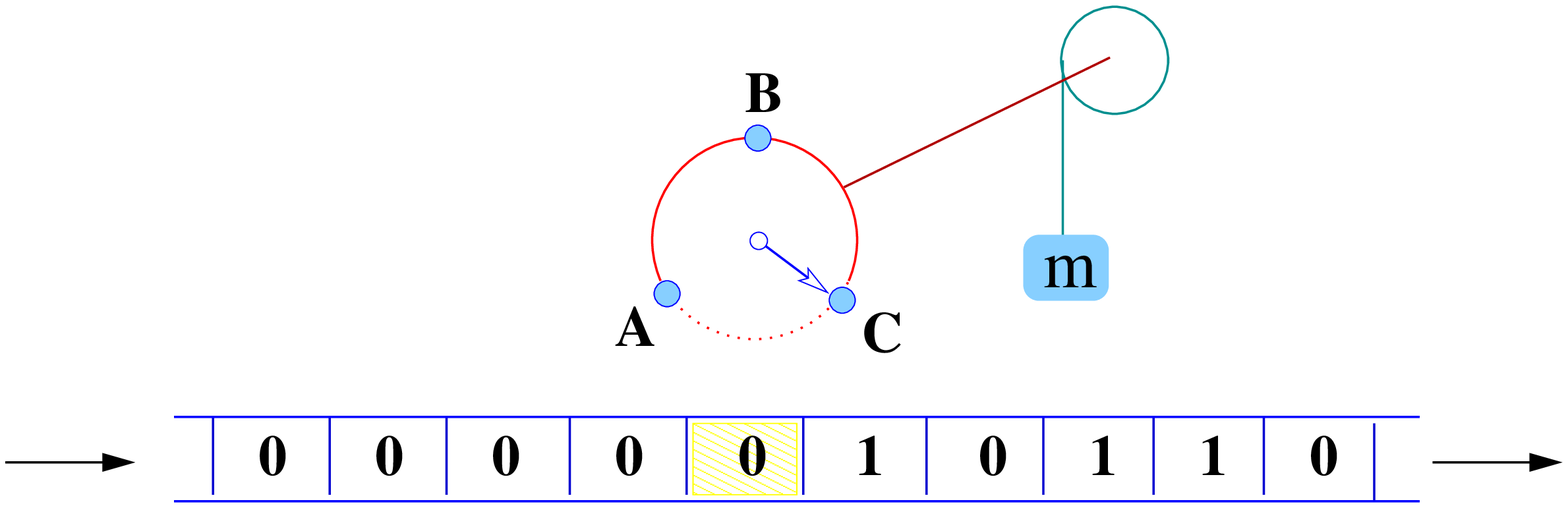}
\caption{A schematic diagram of our model: The demon is a three state system which is coupled with a external load and 
two thermal reservoirs (not shown). A sequence of bits (tape) passes from left to right at a constant speed.
 The nearest bit interact with the demon. For positive load i.e,  $w>$ 0, the mass is lifted at an amount
 $\Delta h$ for every transition C $\rightarrow$ A while for every transition A $\rightarrow$ C the mass
 is lowered by same amount. However  $w <$ 0 the mass is connected to right side of the small circle, so the transition 
  C $\rightarrow$ A lowers the mass and  the transition A $\rightarrow$ C lift it up.}
\label{model1}
\end{center}

\end{figure}
\begin{figure}[!ht]
\vspace{0.5cm}
\begin{center}
 
\includegraphics[width=7cm]{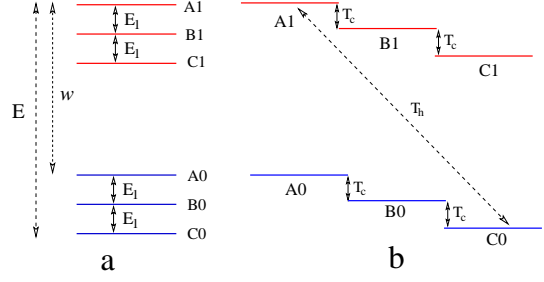}
\caption{Possible six states depending on demon and the bit. (a) the difference between the energy levels 
between A and B is $E_1$ similarly the difference between B and C is also $E_1$. The energy absorbed/released
during any transition between C0 and A1 is denoted as E such that $E=w+2E_1$. (b) All the allowed transitions 
and corresponding bath where energy is exchanged.}
\label{model2}
\end{center}
\end{figure}
\begin{figure}[!ht]
\vspace{0.5cm}
\begin{center}
 
\includegraphics[width=3cm]{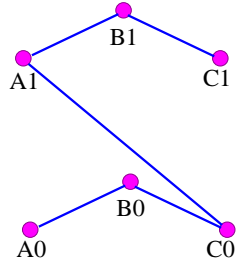}
\caption{All the allowed transitions form a liner chain during each operation interval  $\tau$}
\label{model3}
\end{center}
\end{figure}

All the allowed transitions are shown in Fig.(\ref{model2}) and they form a linear chain Fig.(\ref{model2}).
The weight parameter $\varepsilon$ is defined by
\begin{equation}
 \varepsilon=\tanh \left( \frac{E}{2T_h} \right)
\end{equation}
and is bounded by $-1<\varepsilon<1$.  Let the tape is moving with a speed $\frac{1}{\tau}$ and interaction 
happens  only with the nearest bit (one bit at a time). During time $\tau$, the system (demon) 
evolves along with the bit. Then next bit arrive and earlier bit moves forward containing
the information. Note that only during the interaction with the system the bit may flip.
If $\tau$ is small, system hardy evolves. However, if $\tau$ is large system gets enough opportunity
to evolve along with the bit. Let $p(0)$ and $p(1)$ denotes the probability of 0 and 1 in 
the incoming bit stream. $P(0) + P(1) =1$. Then one can define
\begin{equation}
\delta =p(0)-p(1).
\end{equation}
as the probability of excess number of 0 in incoming bit stream. The Shannon entropy of the incoming
bit stream is defined as
\begin{equation}
 S=-p(0)\ln p(0)-p(1)\ln p(1)
\end{equation}
This Shannon entropy measures amount of disorder present among the bits. However, we have ignored the 
correlation between the successive bits. It also quantifies the amount of information present in the 
bit stream. More disorder is equivalent to more information. Note that, if every bits are zero The Shannon 
entropy is zero. Where it is again true when every states are 1. Shannon entropy becomes maximum when 
both 0 and 1 state are equally probable and its value then become $S_{max}=ln2$. Now if $p'(0)$ and $p'(1)$
represents the probability of 0 and 1 in the outgoing 
bit stream then corresponding Shannon entropy will be
\begin{equation}
 S'=-p'(0)\ln p'(0)-p'(1)\ln p'(1).
\end{equation}
The change of Shannon entropy of the bits, written on the tape, will be simply $\Delta S = S' - S$. 
Now, if   $\Delta S > 0$, it means some information is written on the tape or the bits stream become more 
disordered. On the other if $ \Delta S < 0$, it implies during the interaction with the demon 
some information has been erased and it acts as an erasure.  

If $\phi$ represents average number of clockwise (CW) rotation then
\begin{equation}
 \phi=p'(1)-p(1)=(\delta-\delta')/2.
\end{equation}
Where $\delta’ = p'(0)$ – $p'(1)$. During each CW rotation $E$ energy is absorbed from hot
bath while system performs $w$ amount of work. Hence the average heat dissipated to the hot
bath will be given by
\begin{equation}
 Q_h=-\phi E,
\end{equation}
and average work done on the system will be
\begin{equation}
 W=-\phi w.
\end{equation}
Using the first law on can readily write the average heat dissipated to the cold bath as
\begin{equation}
 Q_c=2\phi E_1.
\end{equation}
Since the system operates in steady state the entropy production of the system (demon) will be zero.
Then total entropy production will be
\begin{equation}
 \Delta S_{tot}=\Delta S + \Delta S_B
\end{equation}
Where $\Delta S_B = \frac{Q_h}{T_h} + \frac{Q_c}{T_c}$ denotes the bath entropy production.  
In next section the model is analyzed taking help of numerical simulation.

\subsection{Results and discussions}

\begin{figure}[!ht]
\vspace{0.5cm}
\begin{center}
 
\includegraphics[width=7cm]{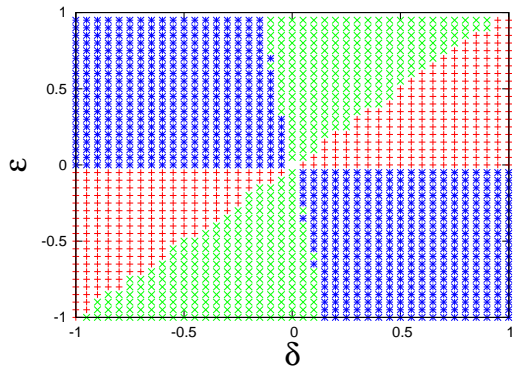}
\caption{Phase diagram when the demon is connected with single bath with  temperature $T_h=1.0$ 
and cycle time $\tau=1.0$. There are thee regions in the phase diagram namely
i) Engine (red plus), ii)  Eraser (green cross), iii) Dud (blue star)
}
\label{mandal}
\end{center}

\end{figure}

This problem may become simpler by taking $E_1=0$ $ (w=E)$ .
This implies  all the three states of the system are degenerate. Due to presence of bits 
the joint state $A1$, $B1$, $C1$ has energy $E$ while $A0$, $B0$, $C0$ has energy zero. The transition 
between $C0$ and $A1$ takes place with the interaction of the bath $T_h$ and $E=w$ energy will be 
exchanged. Then this problem will be reduced to original Mandal Jarzynski model \cite{man12}.  
The temperature of the  bath is set at  $T_h=1.0$. The interaction time with the demon with
each bit is kept fixed at $\tau =1.0$.  Then one can  obtain the phase diagram numerically
as shown in Fig.(\ref{mandal}) by simply varying  $-1< \varepsilon <1$ and  $-1< \delta <1$. 
The system will acts as an engine (red plus region) by extracting heat from single heat
bath and convert it completely into output work.
This is quite surprising. However the entropy of the outgoing bits  will be more 
compared to the incoming bits ($\Delta S >0 $). Hence in the expense of information as a fuel, 
the system performs an engine. In  green region, the system acts as an erasure where it 
erases information ($\Delta S < 0 $) written on the tape while work is done on the system ($W>0$).
In other region in phase space (blue region), it neither perform as an engine ($W>0$) nor 
a erasure ($\Delta S >0 $) and denoted by dud.

Next we move to earlier problem. The temperature of the hot and cold bath is set at $T_h=1.0$ and $T_c=0.5$ 
respectively. The separation between two successive energy levels is taken at 
$E_1 = 0.5$ and $\tau=1.0$.  Then one can obtain the phase 
diagram by simply varying  $-1< \varepsilon <1$ and  $-1< \delta <1$.  It is important to mention that the
system experiences two forces (randomization of bits and pull of gravity) besides the thermal bias between
two baths. The phase diagram is obtained due to the interplay of these three forces.

 \begin{figure}[!ht]
\vspace{0.5cm}
\begin{center}
 
\includegraphics[width=7cm]{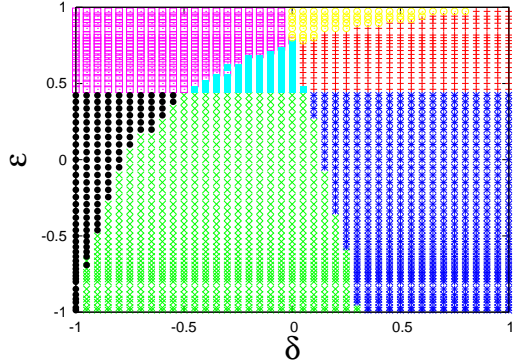}
\caption{Phase diagram when the demon is connected with two baths with  temperature $T_h=1.0$
and $T_c=0.5$. The other parameters set at $E_1$=0.5 and  $\tau=1.0$.
The phase diagram consist of seven regions namely
i)only Engine (red plus), ii) only Eraser (green cross), iii) Dud (blue star), iv) only refrigerator
(pink open box), v) Eraser and Refrigerator( yellow open circle), vi)Erasure and Engine
(cyan full box), vii) Engine and Refrigerator (black full circle)
}
\label{phase-1-0.5}
\end{center}

\end{figure}

The phase diagram (Fig.(\ref{phase-1-0.5})) consists of 7 regions. i) In red plus region the system
performs as an engine by extracting work $(W<0)$ on average while writing information on the tape $(\Delta S>0)$.
The efficiency of an engine    $\eta=\frac{W}{Q_h}$ can exceed Carnot limit ($\eta > 0.5$ 
for this case).  ii) The system acts as an erasure $(\Delta S<0)$ in green cross region, while work is done
in the system $(W>0)$. iii) In pink open box region, it performs as a refrigerator $(Q_c<0)$ by 
transferring heat from cold bath while $W>0$ and $(\Delta S>0)$.

Apart from these three regions, there are another three regions  where combination of any two is possible. 
Such as: iv) in yellow open box region the system simultaneously acts as an erasure $(\Delta S<0)$
and refrigerator $(Q_c<0)$. However, in this regime, work is done on the system on average; v) in the triangular 
area in the middle (cyan full box) thermal bias plays a dominate role in such a way that the demon 
simultaneously performs as an erasure $(\Delta S<0)$ and engine $(W<0)$ by extracting work on average.  
vi) In black full circle region, the demon transfer heat from cold bath $(Q_c<0)$ as well as it extracted
work on average $(W<0)$.  However on average information is written on the tape $(\Delta S>0)$.
The coefficient of performance of the Refrigerator $\sigma=\frac{-Q_c}{W}$ can take value
beyond Carnot limit ($\sigma > 1 $ for this case) in this region and some part of pink box region 
vii) Finally in blue star region, the system does not acts as an engine, erasure or refrigerator and we call it dud.

It is  observed $\Delta S_{tot}>0$ throughout  the phase space. Hence, all the regions are consistent
with the generalized second law of thermodynamics.

\section{Conclusions}
Information and thermodynamics are treated in a single framework. The generalized second law
is proved when system starts and ends at non-equilibrium state.
We have explained  the performance of Szilard engine which seems to violate the second law. 
We have shown that when measurement is performed,  although the energy of the system is not changed,
the non-equilibrium free energy apparently increases.  Hence one can extract energy in cyclic process
from single heat bath using acquired information by measurement.  However, erasure of stored information
requires work to be done on system.  A memory 
device can be used as a information reservoir which can be used to increase the performance of a device.
We have formulated a simple model of autonomous information engine.  We have found the system 
can act as an engine, refrigerator or  an eraser. Even  combination of any two is possible in some
parameter space. We have achieved the  efficiency of the engine to be greater than Carnot limit. The coefficient
of performance  of refrigerator also goes beyond the  Carnot limit. Our findings are consistent with 
the generalized second law of thermodynamics along with information.

\section{Acknowledgement}

One of us (AMJ) thanks DST, India for financial support (through J. C. Bose National Fellow-ship).


\begin{thebibliography}{10}
\bibitem{max71} J. C. Maxwell, Theory of Heat (Longmans, London, 1871).
\bibitem{szi29} L. Szilard, Z. Phys. {\bf 53}, 840 (1929).
\bibitem{lan61} R. Landauer, IBM J. Res. Dev. {\bf 5}, 183 (1961).
\bibitem{esp11} M. Esposito and C. Van Den Broeck \zepl{95}{40004}(2011).
\bibitem{par15}  J. M. R. Parrondo, J. M. Horowitz  and T. Sagawa, Nature Physics {\bf 11}, 131, (2015).
\bibitem{lan91} R. Landauer Phys. Today {\bf 44}(5), 23 (1991).
\bibitem{ran16}S. Rana and A. M. Jayannavar, arXiv 1603.01129 (2016)
\bibitem{man12} D Mandal and C Jarzynski, Proceedings of the National Academy of Sciences, {\bf 109}, 11641, (2012).
\bibitem{man13} D Mandal, H. T. Quan and  C Jarzynski \zprl{111}{030602}{2013}.
\end{thebibliography}
\end{document}